\documentclass[aps,prb,twocolumn,groupedaddress,floats,showpacs,final,floatfix]{revtex4-1}

\usepackage{multirow}

\usepackage{xcolor}
\usepackage{graphicx}
\usepackage{dcolumn}
\usepackage{float}
\usepackage{subfigure}
\usepackage{bm}

\begin{document}
	
\preprint{APS/123-QED}

\title{Quantum-to-classical correspondence in two-dimensional Heisenberg models}

\author{Tao Wang}
\author{Xiansheng Cai}
\affiliation{Department of Physics, University of Massachusetts, Amherst, MA 01003, USA}
\author{Kun Chen}
\affiliation{Department of Physics and Astronomy, Rutgers University, Piscataway, NJ 08854, USA}
\author{Nikolay V. Prokof'ev}
\affiliation{Department of Physics, University of Massachusetts, Amherst, MA 01003, USA}
\affiliation{National Research Center ``Kurchatov Institute," 123182 Moscow, Russia}
\author{Boris V. Svistunov}
\affiliation{Department of Physics, University of Massachusetts, Amherst, MA 01003, USA}
\affiliation{National Research Center ``Kurchatov Institute," 123182 Moscow, Russia}
\affiliation{Wilczek Quantum Center, School of Physics and Astronomy, Shanghai Jiao Tong University, Shanghai 200240, China}

\date{\today}
	
\begin{abstract}
The quantum-to-classical correspondence (QCC) in spin models is a puzzling phenomenon
where the static susceptibility of a quantum system agrees with its classical-system counterpart,
at a different corresponding temperature, within the systematic error at a sub-percent level.
We employ the bold diagrammatic Monte Carlo method to explore the universality of QCC by
considering three different two-dimensional spin-1/2 Heisenberg models. In particular,
we reveal the existence of QCC in two-parametric models.
\end{abstract}
	
\maketitle

\section{Introduction}
\label{sec:intro}

The quantum-to-classical correspondence (QCC) is a recently discovered phenomenon where the static
susceptibility of a certain spin model (at any available temperature $T_Q$ and lattice distance
$\mathbf{r}$) can be accurately reproduced, up to a global normalization factor, by its classical
counterpart at the corresponding temperature $T_C$. The QCC was first revealed by Kulagin {\it et al}.
in Ref.~\onlinecite{kulagin_bold_2013-1} for the square- and triangular-lattice  spin-1/2 Heisenberg
antiferromagnets.\cite{kulagin_bold_2013-1}
QCC was subsequently established for the pyrochlore lattice Heisenberg antiferromagnet in Ref.~\onlinecite{huang_spin-ice_2016}.

\begin{figure}[htb]
\centering
\subfigure{
		\includegraphics[width=0.90\columnwidth]{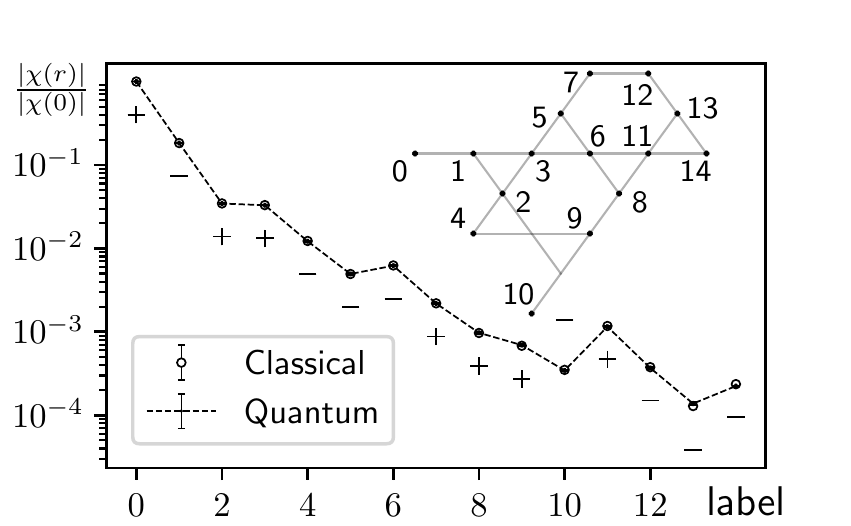}}	
\caption{\label{Fig.Example}
Accurate (within the accuracy bounds) match between the normalized quantum (dots connected by the dashed line) and classical
(open circles) correlation functions
of the kagome-lattice Heisenberg antiferromagnet at \(T_{Q}=1.0\). The sequence of
labeled distances is illustrated in the top right corner.
The sign of the correlation function is indicated explicitly next to each point.
	}
\end{figure}

It is worth noting that the QCC only applies to the the static susceptibility expressed by the correlator
\begin{equation}
\chi(\mathbf{r}) \equiv \int_{0}^{\beta} d\tau \, \chi(\mathbf{r},\tau)=
                        \int_{0}^{\beta} d\tau \, \langle \, \mathbf{S}(0,0) \cdot \mathbf{S}(\mathbf{r},\tau)\,  \rangle \,,
\end{equation}
where $\mathbf{S}(\mathbf{r},\tau)$ is the Matsubara spin-1/2 operator.
The equal-time correlation function, \(\chi(\mathbf{r},\tau=0)\), while having a qualitatively similar spatial profile,
does not match the classical correlation function. It is thus surprising to observe that the
static quantum and classical correlations, despite featuring a highly non-trivial and model-dependent pattern
of sign-alternating spatial fluctuations, demonstrate perfect qualitative and extremely accurate
quantitative agreement (see Fig.~\ref{Fig.Example} and Table~\ref{tab:Example}). Up to now, the origin of QCC still remains unknown, which motivates us to
further study the universal applicability of QCC in two-dimensional (2D) spin systems.

\begin{figure}[htb]
	\centering
	\subfigure[Kagome-lattice model]{
		\label{Fig.lattice.sub.1}
		\includegraphics[width=0.15\textwidth]{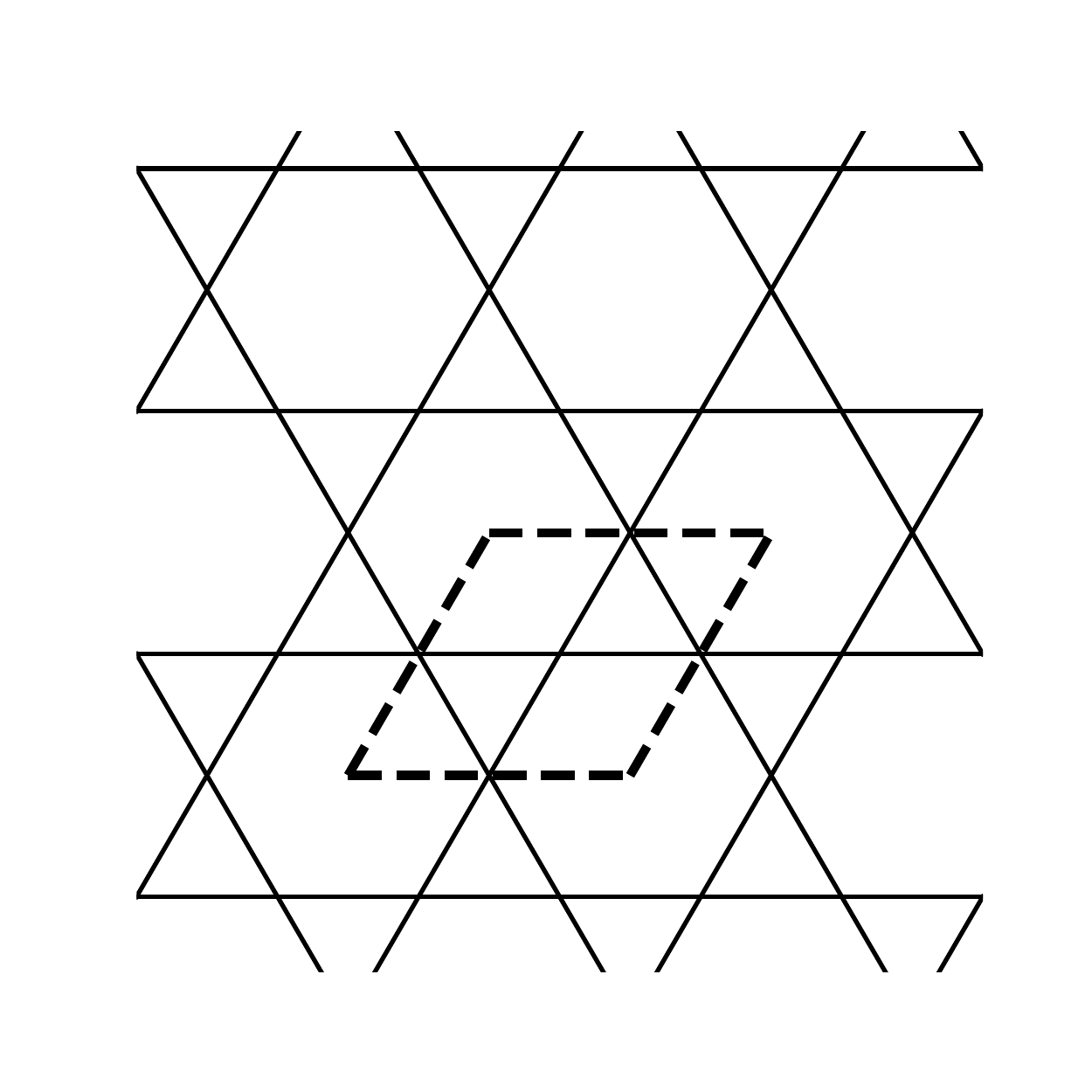}}
	\subfigure[Square-lattice $J_1-J_2$ model]{
		\label{Fig.lattice.sub.2}
		\includegraphics[width=0.15\textwidth]{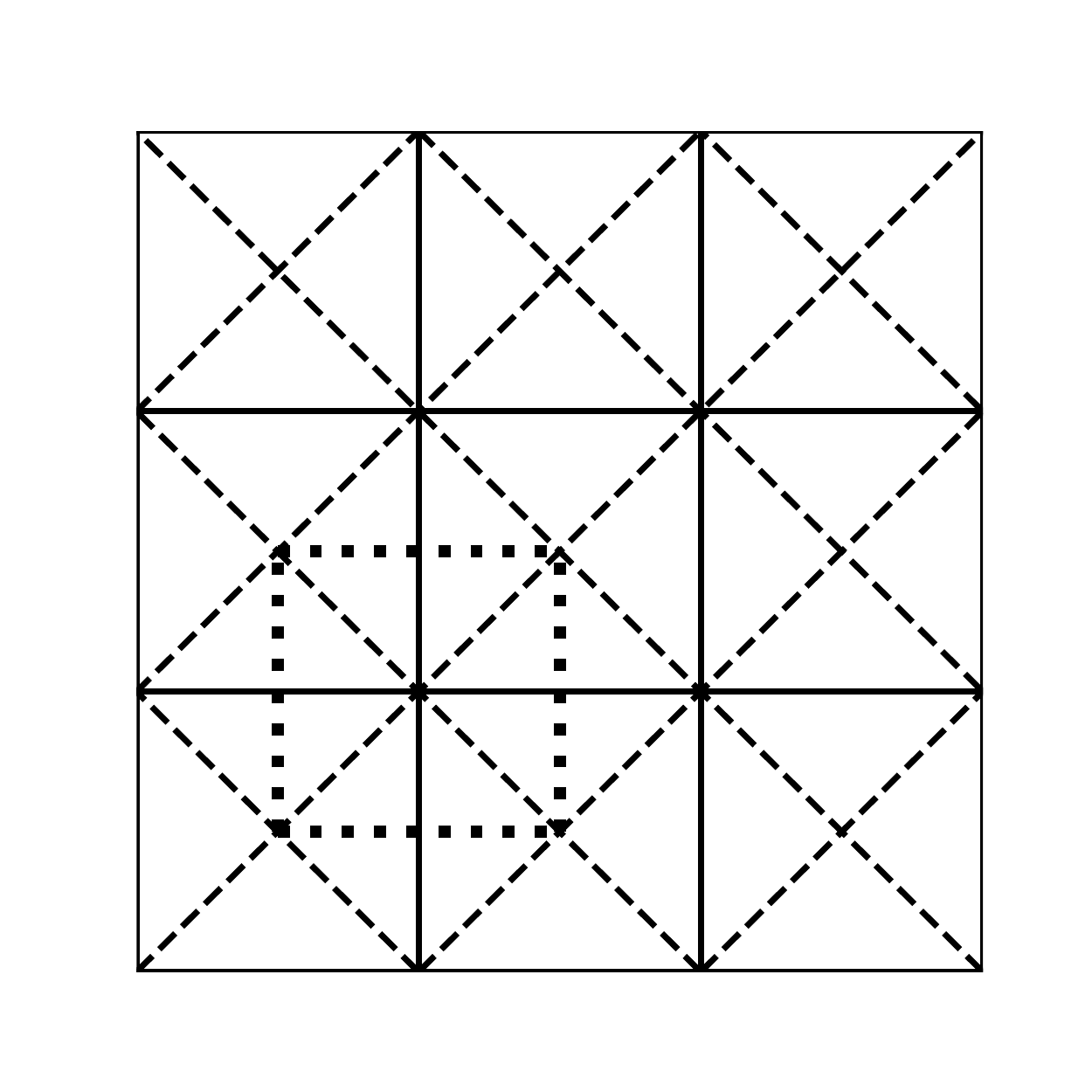}}
	\subfigure[Anisotropic triangular-lattice model]{
		\label{Fig.lattice.sub.3}
		\includegraphics[width=0.15\textwidth]{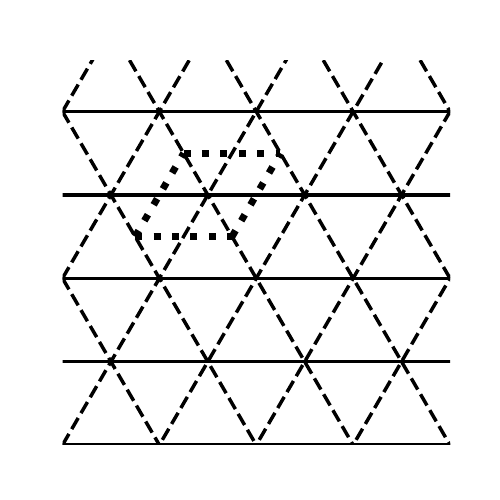}}
	\caption{\label{Fig.lattice}
Specifying interactions and primitive cells for three spin models. In figure \ref{Fig.lattice.sub.1}, all bonds have the same coupling constant J. In figure \ref{Fig.lattice.sub.2} and \ref{Fig.lattice.sub.3}, solid and dashed lines represent
coupling constants $J_1$ and $J_2$ respectively, while dotted lines define the primitive cells.}
\end{figure}

In this article, we verify the existence of the QCC for three 2D frustrated magnets:
the kagome-lattice Heisenberg antiferromagnet (KLHA),
the square-lattice $J_1-J_2$ model,
and the spatially anisotropic triangular-lattice Heisenberg antiferromagnet (ATLHA),
all of which are of great experimental and numerical
interest.\cite{imai_$^63mathrmcu$_2008, mustonen_tuning_2018, coldea_experimental_2001}
All considered Hamiltonians can be described as
\begin{equation}
H \, =\, \sum_{\langle ij \rangle} \, J_{ij} \, \mathbf{S}_{i} \cdot \mathbf{S}_{j} \;,
\end{equation}
where $\langle ij \rangle$ stands for all pairs of interacting lattice sites as illustrated
for each model in Fig.~\ref{Fig.lattice}, and $J_{ij}$ are the corresponding coupling constants.
For KLHA, $J_{ij}=J$, while for the other two models $J_{ij}$ can take two different values,
$J_1$ and $J_2$. The only difference between the quantum and classical models is that
spin-1/2 operators $\mathbf{S}$ are replaced with unit vectors.

%
\begin{table*}[htb]
	\label{tab:Example}
	\begin{center}
		
		\caption{Quantum and classical correlators of the kagome-lattice Heisenberg antiferromagnet and the square-lattice J1-J2 model, 
along with the error bounds on the difference between the two. The error bounds are based on the $3 \sigma$-criterion for purely
statistical Monte Carlo fluctuations and the systematic error of the extrapolation to the infinite diagram order limit for quantum simulations. 
Accurate QCC is observed for all points within the error
bounds.}	
			\begin{ruledtabular}
			\begin{tabular}{ccllllllll}
				
			\multirow{10}{*}{\rotatebox[origin=c]{90}{Kagome}}
			&\textbf{Space Label}&1&2&3&4&5&6&7&8\\
			&Classical Correlator&1.000000&0.182995&0.034589&0.033010&0.012270&0.004932&0.006247&0.002195\\
			&Quantum Correlator&1.000000&0.183641&0.034820&0.033191&0.012339&0.005005&0.006291&0.002225\\
			&Correlator Difference&0.000000&0.000646&0.000231&0.000181&0.000069&0.000073&0.000044&0.000030\\
			& Errorbar&0.000000&0.000978&0.000342&0.000319&0.000148&0.000104&0.000069&0.000052\\
			&\textbf{Space Label}&9&10&11&12&13&14&15&\\
			&Classical Correlator&0.000969&0.000681&0.000350&0.001175&0.000377&0.000129&0.000236&\\
			&Quantum Correlator&0.000994&0.000703&0.000353&0.001174&0.000378&0.000141&0.000229&\\
			&Correlator Difference&0.000025&0.000022&0.000003&0.000001&0.000001&0.000012&0.000007&\\
			&Errorbar&0.000029&0.000024&0.000013&0.000025&0.000016&0.000012&0.000012&\\
			\toprule
			\multirow{10}{*}{\rotatebox[origin=c]{90}{Square}}
			&\textbf{Space Label}&1&2&3&4&5&6&7&8\\
			&Classical Correlator&1.000000&0.166453&0.042818&0.038162&0.017860&0.001943&0.009799&0.005713\\
			&Quantum Correlator&1.000000&0.166971&0.042386&0.038169&0.017840&0.002026&0.009810&0.005695\\
			&Correlator Difference&0.000000&0.000518&0.000432&0.000007&0.000020&0.000083&0.000011&0.000018\\
			& Errorbar&0.000000&0.002629&0.000448&0.000914&0.000219&0.000217&0.000299&0.000140\\
			&\textbf{Space Label}&9&10&11&12&13&14&15&\\
			&Classical Correlator&0.000428&0.000320&0.002663&0.001741&0.000347&0.000073&0.000019&\\
			&Quantum Correlator&0.000394&0.000327&0.002653&0.001731&0.000340&0.000078&0.000003&\\
			&Correlator Difference&0.000034&0.000007&0.000010&0.000010&0.000007&0.000005&0.000016&\\
			&Errorbar&0.000069&0.000046&0.000123&0.000083&0.000040&0.000046&0.000037&\\
			\end{tabular}
			\end{ruledtabular}
	\end{center}
\end{table*}

On the one hand, it is believed that the quantum KLHA is one of the most promising candidates
for a spin liquid ground state that does not break the spin-rotation and lattice-translation symmetries.\cite{sachdev_kagomeifmmodeacuteelsetextasciiacutefi-_1992,yan_SpinLiquidGroundState_2011,depenbrock_NatureSpinLiquidGround_2012,iqbal_GaplessSpinliquidPhase_2013}
On the other hand, it has been reported that the classical KLHA is located at a tricritical point
where three different ordered states coexist.\cite{messioKagomeAntiferromagnetChiral2012}
The proposed quantum and classical ground states are, thus, dramatically different,
which apparently denies the existence of QCC at least at low enough temperature.
We verify that the QCC remains valid at temperatures $T/J \geq 1/3$.
Unfortunately, limitations of the bold diagrammatic Monte Carlo method (BDMC) based on the $G^2W$-expansion
\cite{kulagin_bold_2013-1} do not allow us to access lower temperatures
to ensure that the ground-state properties are dominating in the correlation
function.\cite{shimokawa_finite-temperature_2016}
Whether QCC is valid at much lower temperature remains to be seen in the future.

The square-lattice $J_1-J_2$ model enables us to explicitly check the validity of QCC
in the different phases of the same system.
Numerous previous work has established the rich ground-state phase diagram of this model
with respect to changing the $J_{2}/J_{1}$ ratio.\cite{mustonen_tuning_2018}
Apart from the spin liquid state predicted for \(0.41\leq J_{2}/J_{1}\leq0.62\),\cite{jiang_spin_2012}
it also features three ordered states: ferromagnetic (FM), N\'{e}el antiferromagnetic (NAF),
and collinear antiferromagnetic (CAF). We choose the following parameter sets in this work:
$(J_1=1.0, \, J_2=0.5)$ to address the mostly frustrated case and
$(J_1=-1.0, \,J_2=0.4)$ in the CAF phase (notice the ferromagnetic sign of the nearest neighbor interaction).
Here and in what follows, we choose the modulus of $J_{1}$ as the unit of energy.

The ATLHA model is chosen specifically to study how moderate anisotropy in the coupling constants effects
the QCC. In this case, we choose \(J_{2}/J_{1}=0.33\), which is the same as the ratio used to explain
experimental data in \(Cs_{2}CuCl_{4}\).\cite{coldea_experimental_2001}
When the anisotropy is very strong, the ATLHA model resembles decoupled 1D chains, for which the QCC does not hold.\cite{kulagin_bold_2013-1}
It appears that observing the crossover between the 1D and 2D behavior requires very small
ratios of the coupling constants, and the fascinating QCC phenomenon is robust against anisotropy.
\begin{figure}[htb]
	\centering	
	\subfigure{
		\label{Fig.all_QCC.Cor.2}
		\includegraphics[width=0.90\columnwidth]{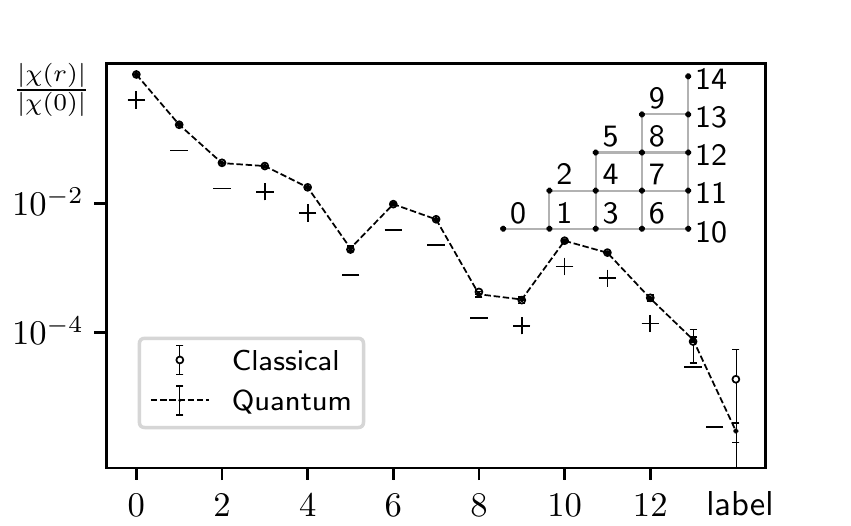}}	
	\caption{\label{Fig.all_QCC}
Accurate match of the normalized static quantum (dots connected by the dashed line) and classical (open circles)
correlation functions for the square-lattice $J_1-J_2$ model at \(T_{Q}=1.0\).
The sequence of labeled distances is illustrated in the top right corner.
The sign of the correlation function is indicated explicitly next to each point.
	}
\end{figure}

To obtain the static spin-spin correlation function for quantum models we employ the  BDMC method
that allows one to study any frustrated spin model in the cooperative paramagnetic regime at temperatures
below the exchange coupling constant $J$.\cite{kulagin_bold_2013-1,huang_spin-ice_2016}
The relative accuracy of the converged BDMC results is  $\sim$1\% (the loss of convergence
is the prime reason preventing the method from being used at very low temperature).
All models were simulated on lattices with periodic boundary conditions and system sizes 
$L\times L=16 \times 16$ in terms of primitive cells. These system sizes are
much larger than the correlation length to ensure that finite-size corrections to presented results 
are negligible (the correlation functions decrease by about four orders of 
magnitude before reaching distances $L/4$ along the primitive cell directions).
The primitive cells and sample geometry are showed in Fig.\ref{Fig.lattice}.

Establishing QCC for single-parameter models boils down to one-to-one correspondence between the temperatures
of quantum, $T_Q$, and classical, $T_C$, systems, for which the difference between the normalized
correlation functions, $\chi(r)/\chi(0)$, is minimized. This ``one-dimensional" $T_Q$-to-$T_C$ mapping
applies to KLHA. More interesting results are obtained for the other two models, both of which feature
an additional model parameter \(J_{2}\). It turns out that not only the temperature but also $J_2$
need to be fine-tuned to obtain the best match between the quantum and classical correlation functions
if we choose to stay in the same model subspace.
To be more specific, we find that for the quantum model with $J_{2}^{Q}\neq 0$ at temperature
$T_{Q}$, the matching classical counterpart should be taken with $J_2^C \neq J_{2}^{Q}$ at temperature $T_C$
(asymptotically, \( J_2^C \to J_{2}^{Q}\) at high temperature). This constitutes a ``two-dimensional"
($T_Q$,$J_2^Q$)-($T_C$,$J_2^C$) mapping.

In what follows, we establish that at all accessible temperatures
all models demonstrate a perfect (within error bars) match between
the static quantum and classical correlation functions. We discuss properties
of the correspondence mapping, and conclude with broader implications of this work,
as well as perspectives for future developments.

\section{Results}
\label{sec:results}

The precise protocol for establishing the QCC is as follows.
We first compute the static correlation function of the quantum
system by the BDMC method. The answer for its classical counterpart \(\chi_{C}(\mathbf{r})\)
was obtained by the conventional single-spin flip Monte Carlo method.
Next, we normalize the quantum result to unity at the origin
[\(\chi_{C}(\mathbf{r}=0)=1\) automatically], to obtain \(f(\mathbf{r})=\chi(\mathbf{r})/\chi(0)\).
Finally, we fine-tune classical system parameters---which
are, in our case, \(T_{C}/J_1\) and \(J_2^C\)---to find the best fit to
the \(f(\mathbf{r})\) curves. We repeat this process at different temperatures
\(T_{Q}\) or values of $J_2^{Q}$, to obtain the correspondence curves.

Note that we have only one or two fitting parameters to described the entire
functional dependence of $f$ on distance, including numerous, and often irregular,
sign changes and an order of magnitude strong fluctuations. Remarkably, all these
features can be reproduced by the classical model at all distances within the error
bounds of our calculations (often at the sub-percent level for several closets sites).
In Figs.~\ref{Fig.Example} and \ref{Fig.all_QCC}, we show examples of QCC
for KLHA and the square-lattice $J_1-J_2$ model at \(T_{Q}=1.0\).
The results shown in both plots are also presented in Table~\ref{tab:Example}
because for most points the errorbars are smaller than point sizes.
We observe that an accurate match can be achieved, and this holds at all temperatures
accessible to us and for all models studied in this work. As of now, no exception from
the QCC ``rule" was found in dimensions $d>1$.

\begin{figure}[!t]
	\centering
	\subfigure{
		\label{Fig.all_TT.sub.1}
		\includegraphics[width=0.80\columnwidth]{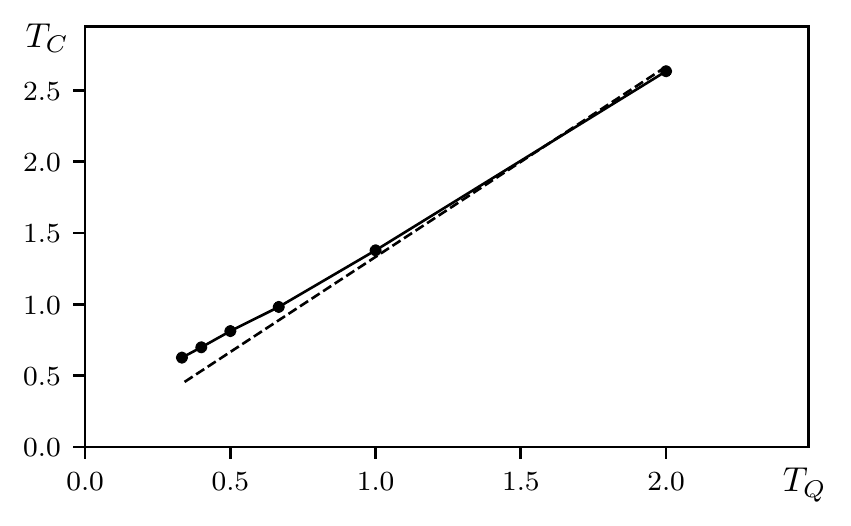}}
	\subfigure{
		\label{Fig.all_TT.sub.2}
		\includegraphics[width=0.80\columnwidth]{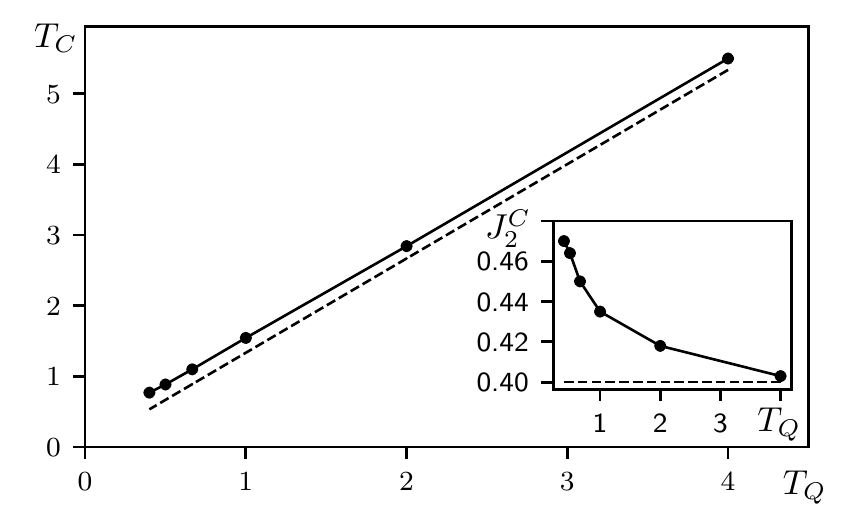}}
	\subfigure{
		\label{Fig.all_TT.sub.3}
		\includegraphics[width=0.80\columnwidth]{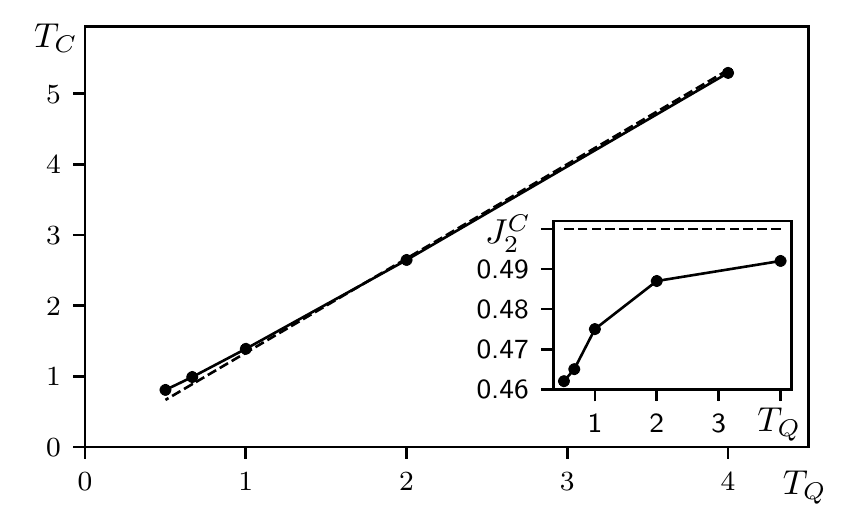}}
	\subfigure{
		\label{Fig.all_TT.sub.4}
		\includegraphics[width=0.80\columnwidth]{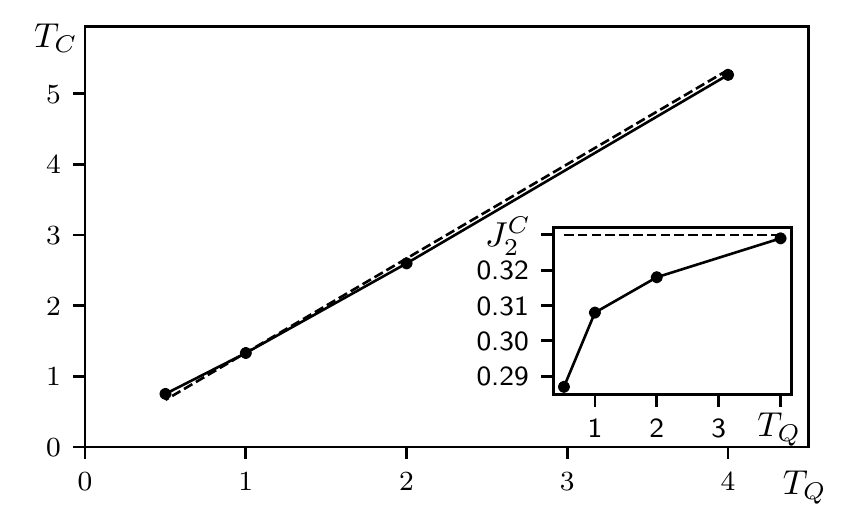}}
	\caption{\label{Fig.all_TT}
		Correspondence curves of all models. From top to the bottom: KLHA; square-lattice CAF ($J_1=-1.0$, $J_2=0.4$ );
		square-lattice QSL ($J_2/J_1=0.5$); ATLHA ($J_2/J_1=0.33$). The high-temperature asymptotic relations \(T_{C}=(4/3)T_{Q}\) and \(J_{2}^{C}=J_{2}^{Q}\) are indicated by the dashed lines.
	}
\end{figure}

The free parameters of the classical model, \(T_{C}\) and \(J_{2}^{C}\)
are plotted in Fig.~\ref{Fig.all_TT} as a function of the quantum model temperature \(T_{Q}\),
together with the high-temperature asymptotic relations \(T_{C}=(4/3)T_{Q}\) and \(J_{2}^{C}=J_{2}^{Q}\).
(For models with two parameters, the QCC represents a 2D mapping. If we keep $J_{2}^{Q}$ fixed, we can still
present it as the correspondence curves.)
It is worth noting that \(J_{2}^{C}\) of the square-lattice $J_1-J_2$ model approaches \(J_{2}^{Q}\)
from different sides when we change the sign of $J_1$. Mapping of spin-spin correlation functions between the
quantum and classical models is rather standard and expected in two limiting cases. At \(T/J\gg 1\), it can be established
analytically by looking at the lowest-order high-temperature series expansion contribution capturing
the weak short-range correlations. At distances beyond the small correlation length, both systems are
described by the universal coarse-grained field statistics.
The QCC in the cooperative paramagnetic regime, $T/J \lesssim 1$,
is fundamentally different from these limiting cases: on the one hand, correlations at short distance
are strong and far from being accurately described by the lowest-order high-temperature series expansion,
on the other hand, the correlation length remains short and the coarse-grained description is not applicable.

\section{Discussion}
\label{sec:dis}
Using the BDMC technique we computed the static spin-spin correlations as functions of distance
for three different frustrated spin models, including the cooperative paramagnetic regime that,
as far as we know, cannot be addressed for large system sizes by any of the other numerical methods.
We found that all systems feature the non-trivial quantum-to-classical correspondence.
We measured the correspondence curves for each model down to temperatures below the exchange coupling
constant and verified that each curve follows the expected asymptotic behavior in the high-temperature limit.

Future numerical work with respect to QCC can follow two different routes.
(i) Extend the low-temperature range for quantum systems. Our current implementation of the
BDMC technique faces convergence problems at temperature $T\ll J$ and does not allow us to
obtain data at sufficiently low $T$ for reliable extrapolation to the ground state. Making predictions
based on QCC with regards to the spin liquid ground state is not possible under these conditions.
There exist numerous alternative formulations of the diagrammatic expansion \cite{shiftedaction}
and ways of regrouping and re-summing diagrammatic series; some of them may prove helpful in extending
the range of temperatures where the diagrammatic Monte Carlo technique works.
(ii) Expand the ``family" of models demonstrating the QCC in dimensions $d>1$,
or find exceptions from the ``rule." Without proper theoretical understanding of its origin,
it is worth exploring how other model features, such as long-range coupling, effect QCC.

\begin{acknowledgements}

 This work was supported by the Simons Collaboration on the Many Electron Problem, the National Science Foundation under the grant DMR-1720465, and the MURI Program ``New Quantum Phases of Matter" from AFOSR.

\end{acknowledgements}

\end{document}